 \documentclass[smallabstract,smallcaptions]{dccpaper}

\usepackage{epsfig}
\usepackage{citesort}
\usepackage{amsmath}
\usepackage{amssymb}
\usepackage{color}
\usepackage{url}
\usepackage[colorlinks=true,linkcolor=blue]{hyperref}%
\usepackage{cite}
\usepackage[caption=false,font=footnotesize,labelfont=rm,textfont=rm]{subfig}
\newlength{\figurewidth}
\newlength{\smallfigurewidth}
\usepackage{bm}

\begin{document}

\title
{\large

\textbf{Towards Efficient 3D Gaussian Human Avatar Compression: A Prior-Guided Framework}
}

\author{%
Shanzhi Yin$^{1}$, Bolin Chen$^{2,3,4}$, Xinju Wu$^{1}$, Ru-Ling Liao$^{2}$, Jie Chen$^{2,3}$,\\ Shiqi Wang$^{1}$ and Yan Ye$^{2}$\\[0.5em]
{\small\begin{minipage}{\linewidth}\begin{center}
\begin{tabular}{ccc}
$^{1}$City University of Hong Kong  & $^{2}$DAMO Academy, Alibaba Group  \\ $^{3}$ HuPan Laboratory & $^{4}$ Fudan University
\end{tabular}
\end{center}\end{minipage}}
}
\maketitle
\thispagestyle{empty}
\vspace{0.1cm}
\begin{abstract}
This paper proposes an efficient 3D avatar coding framework that leverages compact human priors and canonical-to-target transformation to enable high-quality 3D human avatar video compression at ultra-low bit rates. The framework begins by training a canonical Gaussian avatar using articulated splatting in a network-free manner, which serves as the foundation for avatar appearance modeling. Simultaneously, a human-prior template is employed to capture temporal body movements through compact parametric representations. This decomposition of appearance and temporal evolution minimizes redundancy, enabling efficient compression: the canonical avatar is shared across the sequence, requiring compression only once, while the temporal parameters, consisting of just 94 parameters per frame, are transmitted with minimal bit-rate.
For each frame, the target human avatar is generated by deforming canonical avatar via Linear Blend Skinning transformation, facilitating temporal-coherent video reconstruction and novel view synthesis. 
Experimental results demonstrate that the proposed method significantly outperforms conventional 2D/3D codecs and existing learnable dynamic 3D Gaussian splatting compression method in terms of rate-distortion performance on mainstream multi-view human video datasets, paving the way for seamless immersive multimedia experiences in meta-verse applications.
\end{abstract}




\Section{Introduction}
The emerging immersive multi-media applications like meta-verse and mixed-reality demand efficient storage and transmission of human-centered volumetric videos.
Unlike 2D human videos~\cite{mttf}, volumetric videos integrate temporal scene dynamics, depth information, and multi-viewpoint perspectives, resulting in an exponential increase in data volume. Meanwhile, the modeling of 3D human avatars can be realized by diverse formats, such as mesh or point cloud, which is not compatible with mainstream video coding standards like High Efficiency Video Coding~(HEVC)~\cite{hevc} and Versatile Video Coding~(VVC)~\cite{vvc}.
Furthermore, recent advancements in 3D vision technologies have shifted the paradigm of human avatar modeling from traditional graphics-based methods, such as Shape Completion and Animation for PEople~(SCAPE)~\cite{scape} and Skinned Multi-Person Linear model~(SMPL)~\cite{smpl}, to neural-based approaches. Notably, implicit neural representations, such as occupancy fields~\cite{pifu} and Neural Radiance Fields (NeRF)~\cite{humannerf}, have become prevalent for modeling volumetric videos. However, the Multi-Layer Perceptron (MLP) architectures employed in these methods often result in a large number of network parameters, leading to high storage demands and substantial computational costs for training and rendering~\cite{3d-avatar-survey}. 

In contrast, 3D Gaussian Splatting (3DGS)~\cite{3dgs} explicitly optimizes the attributes of 3D Gaussians and employs splatted projection with $\alpha$-blending for rendering, providing a more efficient alternative to implicit methods. Consequently, a growing body of work has demonstrated the effectiveness of 3DGS for human avatar modeling. For instance, Animatable 3D Gaussian~\cite{Animatable-3d-gaussian} maps sampled points on a skinned 3DGS human to a colored canonical space, then deforms the canonical avatar to a posed space using rigid transformations derived from target bone configurations. Similarly, 3D-GS Avatar~\cite{3dgs-avatar} integrates both rigid and non-rigid transformations with view-dependent color mapping, while GauHuman~\cite{gauhuman} achieves real-time rendering through linear blend skinning (LBS) pose transformations and network-based refinements. GaussianAvatar~\cite{gaussianavatar} further enhances rendering quality by predicting 3DGS attributes from pose and appearance features.
However, these approaches do not address the compression of 3DGS-based human avatars, which requires both efficient appearance storage and compact temporal representation. In parallel, recent advances in 3DGS compression~\cite{3d-avatar-survey,gscodec} have shown promising results in reducing storage requirements for 3DGS scenes through techniques such as pruning~\cite{Lp-3dgs}, inter-gaussian prediction~\cite{scaffold}, rate-distortion optimization~\cite{compgs}, vector quantization~\cite{hac}, and 3D-to-2D projection~\cite{gs-sort}. Nevertheless, these methods are not tailored to human avatars and fail to leverage domain-specific priors to enhance compression efficiency.

In this paper, we make the first attempt to propose an efficient human-prior-guided 3D Gaussian avatar compression framework that enables ultra-low bit-rate transmission of 3DGS avatar videos while achieving high-quality reconstruction and novel-view rendering.
In particular, network-free 3D Gaussian avatar representation is adopted to rely solely on gaussian attributes for appearance modeling. It eliminates the need for identity-dependent calibration or refinement networks, thereby reducing additional bit-rate consumption. Furthermore, a parametric human model serves as the human-prior to enable efficient temporal representations, allowing avatar body movements to be controlled with highly compact parameters.
Finally, a canonical-to-target transformation with articulated Gaussian splatting employs Linear Blend Skinning (LBS) transformations to derive target avatars and provides a unified canonical avatar that can be efficiently compressed using off-the-shelf 3DGS codecs.
The main contributions of this paper are summarized as follows,
\begin{itemize}
\vspace{-0.2cm}
\item{We design a domain-specific 3DGS compression framework for human avatar, which decompose volumetric human videos into canonical 3DGS avatars and compact temporal human-prior parameters.} 
\vspace{-0.2cm}
\item{We develop network-free canonical 3DGS avatar representations and employ LBS transform to derive target 3DGS avatar, significantly enhancing the efficiency of both appearance and temporal representations.} 
\vspace{-0.2cm}
\item{We compare the proposed method against both conventional 2D/3D codecs as well as learning-based dynamic 3DGS compression method on ZJU-MoCap~\cite{nerualbody} and MonoCap~\cite{monocap_source} datasets, demonstrating superior rate-distortion performances and high-quality multi-view rendering. } 
 
\end{itemize}

\Section{The Proposed Efficient 3DGS Human Avatar Coding Framework}
\SubSection{2.1 \ Overall Framework}
\begin{figure*}[t]
\centering
\centerline{\includegraphics[width=15cm]{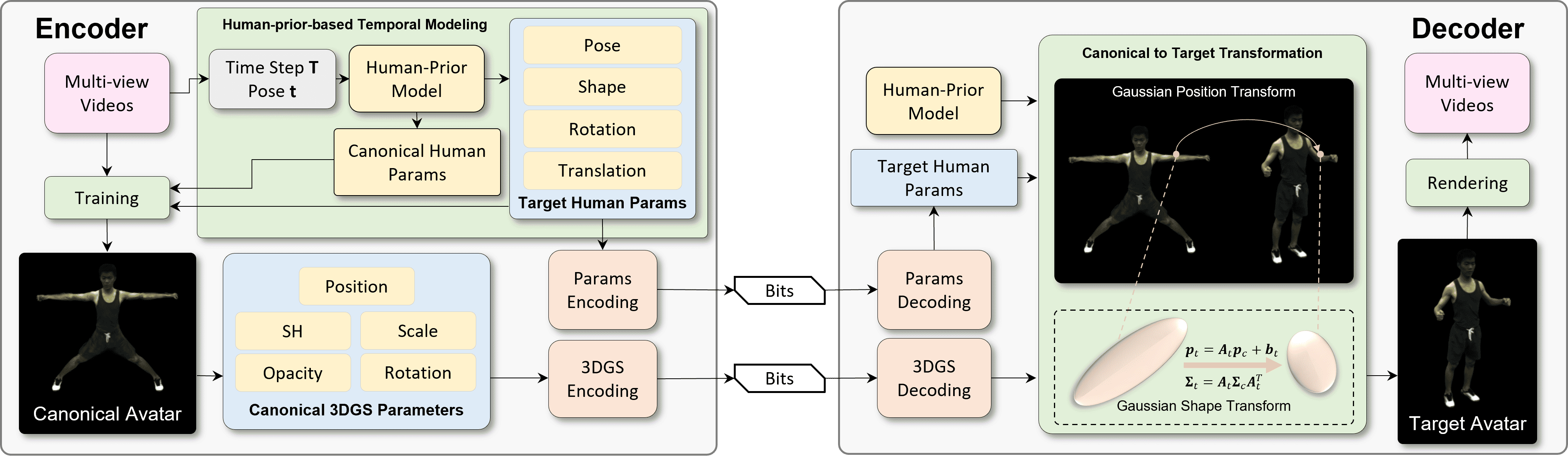}}

\caption{The detailed structure of proposed human-prior-guided efficient 3D gaussian human avatar compression framework.} 

\label{framework_detail}
\end{figure*}

The detailed structure of proposed framework is shown in Fig.~\ref{framework_detail}. At the encoder side, a canonical 3D gaussian avatar is trained with multi-view videos. Following~\cite{gauhuman}, the 3D gaussian is initialized with human-prior-based vertices instead of Structure-from-Motion~\cite{sfm}, which can accelerate convergence and improve rendering quality. The trained canonical avatar is shared across all frames in the avatar sequence, and adopts a ``star-shaped pose", which is characterized by maximally extended limbs, with arms and legs spread outward to form a symmetrical, star-like silhouette. The canonical avatar is then represented as a standard 3DGS with attributes including position, scale, rotation, opacity and spherical harmonics~(SH), which can be compressed by off-the-shelf conventional codecs~\cite{gsc-anchor}. To model temporal body movements, a human-prior model such as SMPL~\cite{smpl} or SMPL-X~\cite{smplx} is utilized to extract target human parameters at each timestamp, including pose, shape, rotation, and translation. Finally, the human parameters are coded by arithmetic coding such as Context Adaptive Binary Arithmetic Coding~(CABAC). 

At the decoder side, the canonical 3DGS avatar and target human parameters are decoded by conventional codecs and arithmetic decoding, respectively. Then, the transformation matrices and translation vectors can be derived from target human parameters by combining the corresponding parameters from each skeleton joints. Accordingly, canonical-to-target transformation is then performed to deform the canonical avatar to the target avatar. Specifically, the position of each 3DGS is transformed using LBS algorithm and each 3D Gaussian is reshaped by adjusting its covariance matrix based on the rotation matrix derived from the LBS transformations.
Finally, the target avatar is rendered to reconstruct multi-view videos as well as enable novel-view synthesis.

\SubSection{2.2 \ Human-prior-based Temporal Modeling}
To track temporal pose changes of the human avatar, a human-prior model is employed to define canonical human parameters and extract target human parameters from each frame. We denote the human-prior model as $M$, and the pose parameter and shape parameter of canonical and target human are represented as $\bm{\theta}_{c}, \bm{\theta}_{t}$ and $\bm{\beta}_{c}, \bm{\beta}_{t}$, respectively. The canonical human is then defined as:
\begin{equation}
     \mathbf{p}_c, \mathbf{J}_c = M(\bm{\theta}_c, \bm{\beta}_c),
\end{equation}
where $\mathbf{p}_c$ represents the canonical vertex positions and $\mathbf{J}_c$ denotes the corresponding joint locations. Similarly, the target human can be obtained by
\begin{equation}
     \mathbf{p}_t, \mathbf{J}_t = M(\bm{\theta}_t, \bm{\beta}_t),
\end{equation}
where $\mathbf{p}_t$ represents the target vertex positions and $\mathbf{J}_t$ denotes the corresponding joint locations. To get the world-coordinate-based vertex positions $\overline{\mathbf{p}}_t$, the global rotation matrix $\textbf{R}_{t}$ and translation vector $\textbf{T}_{t}$ should be applied as,
\begin{equation}
\label{world_trans}
\overline{\mathbf{p}}_t = \mathbf{p}_t\textbf{R}_{t}^{T} + \textbf{T}_{t}.
\end{equation}

By sharing a unified canonical pose and shape across all avatar sequences, only the target pose parameters $\bm{\theta}_t$, shape parameters $\bm{\beta}_t$, and global rotation $\mathbf{R}_t$ and translation $\mathbf{T}_t$ need to be transmitted at the encoder side, significantly reducing temporal redundancy for avatar movements and enabling ultra-low bit-rate compression of human avatar sequences. In practice, SMPL~\cite{smpl} is employed as the human-prior model, utilizing 72 pose parameters ($\bm{\theta}_{t}$), 10 shape parameters ($\bm{\beta}_{t}$), a $3\times3$ global rotation matrix ($\mathbf{R}_{t}$), and a $1\times3$ global translation vector ($\mathbf{T}_{t}$), resulting in a total of 94 parameters per frame, which can be decoded by arithmetic coding and transmitted to decoder side.

\SubSection{2.3 \ Canonical-to-Target Transformation}
To fully leverage human-prior representations and accurately recover the target avatar in each frame, inspired by GauHuman~\cite{gauhuman}, an LBS-based canonical-to-target transformation is employed to deform both the positions and shapes of gaussians, which can be derived from pose and shape parameters of human-prior model. Specifically, at the decoder side, the target human parameters are decoded as $\hat{\bm{\theta}}_{t}$, $\hat{\bm{\beta}}_{t}$, $\hat{\bm{R}}_{t}$, $\hat{\bm{T}}_{t}$, and $\hat{\bm{J}}_{t}$ can be derived by human prior model. Then, the translation matrix $\textbf{A}$ from canonical to target human can be given by,
\begin{equation}
    \textbf{A}(\hat{\bm{J}}_{t}, \hat{\bm{\theta}}_{t}) = \sum_{i=1}^{K}\omega_{k}\textbf{A}_{k}(\hat{\bm{J}}_{t}, \hat{\bm{\theta}}_{t}), 
\end{equation}
where $w_{k}$ denotes the LBS weight of the $k$th joint and $\textbf{A}_{k}$ denotes the rotation matrix of the $k$th joint. Similarly, the translation vector $\textbf{b}$ between canonical and target human can be given by,
\begin{equation}
    \textbf{b}(\hat{\bm{J}}_{t}, \hat{\bm{\theta}}_{t}, \hat{\bm{\beta}}_{t}) = \sum_{i=1}^{K}\omega_{k}\textbf{b}_{k}(\hat{\bm{J}}_{t}, \hat{\bm{\theta}}_{t}, \hat{\bm{\beta}}_{t}),
\end{equation}
where $\textbf{b}_{k}$ denotes the translation matrix of the $k$th joint.
Subsequently, the position of target gaussians can be estimated by the vertices transform under the human-prior model~\cite{smpl},
\begin{equation}
    \hat{\textbf{p}}_{t} = \textbf{A}(\hat{\bm{J}}_{t}, \hat{\bm{\theta}}_{t}) \textbf{p}_{c} + \textbf{b}(\hat{\bm{J}}_{t}, \hat{\bm{\theta}}_{t}, \hat{\bm{\beta}}_{t}),
\end{equation}
Then, using equation(\ref{world_trans}), the estimated target gaussian position can be further transformed to world coordinate with,
\begin{equation}
\label{target_pos}
\overline{\hat{\mathbf{p}}}_t = \hat{\textbf{p}}_{t}\hat{\bm{R}}_{t}^{T} + \hat{\bm{T}}_{t}.    
\end{equation}
Meanwhile, the shape of gaussians are adjusted via their covariance,
\begin{equation}
\label{target_cov}
    \bm{\Sigma}_{t} =\textbf{A}(\hat{\bm{J}}_{t}, \hat{\bm{\theta}}_{t})\bm{\Sigma}_{c}\textbf{A}(\hat{\bm{J}}_{t}, \hat{\bm{\theta}}_{t})^{T} = \textbf{A}(\hat{\bm{J}}_{t}, \hat{\bm{\theta}}_{t})\textbf{R}_{c}\textbf{S}_{c}\textbf{S}_{c}^{T}\textbf{R}_{c}^{T}\textbf{A}(\hat{\bm{J}}_{t}, \hat{\bm{\theta}}_{t})^{T},
\end{equation}
where $\bm{\Sigma}_{t}$ denotes the covariance of target gaussian, and $\textbf{R}_{c}$  and $\textbf{S}_{c}$ denotes the rotation and scale of canonical gaussian, respectively. By utilizing canonical-to-target transformation at the decoder side, the target 3DGS avatar is reconstructed using highly compact human pose and shape parameters alongside the canonical 3DGS avatar, facilitating efficient multi-view video reconstruction and high-quality novel-view synthesis.

\begin{figure}[t!]
\centering
\subfloat[ZJU-MoCap dataset]{\includegraphics[width=1 \textwidth]{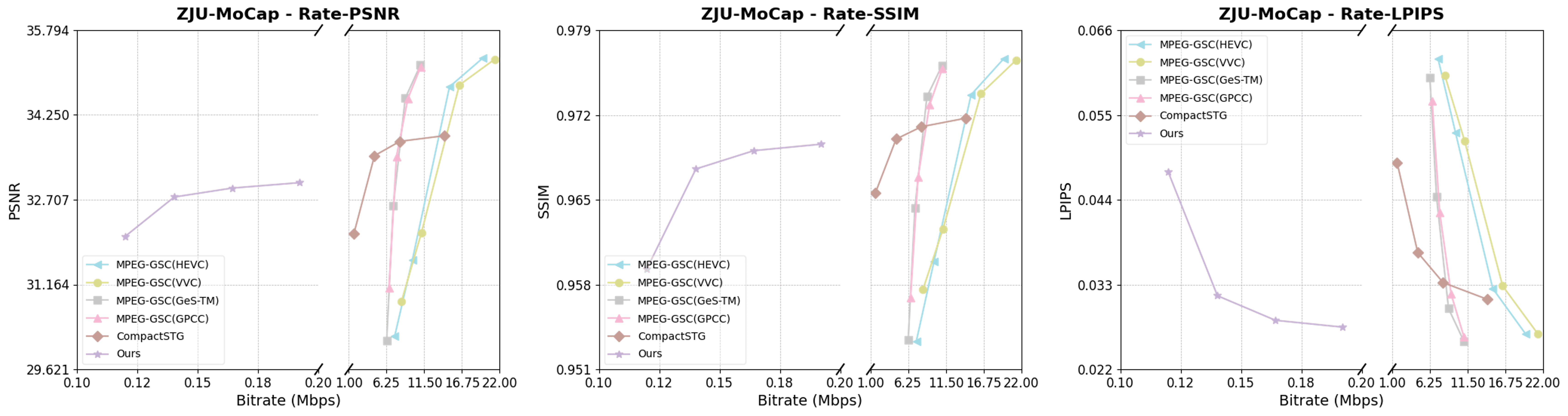}}
\\
\subfloat[MonoCap dataset]{\includegraphics[width=1 \textwidth]{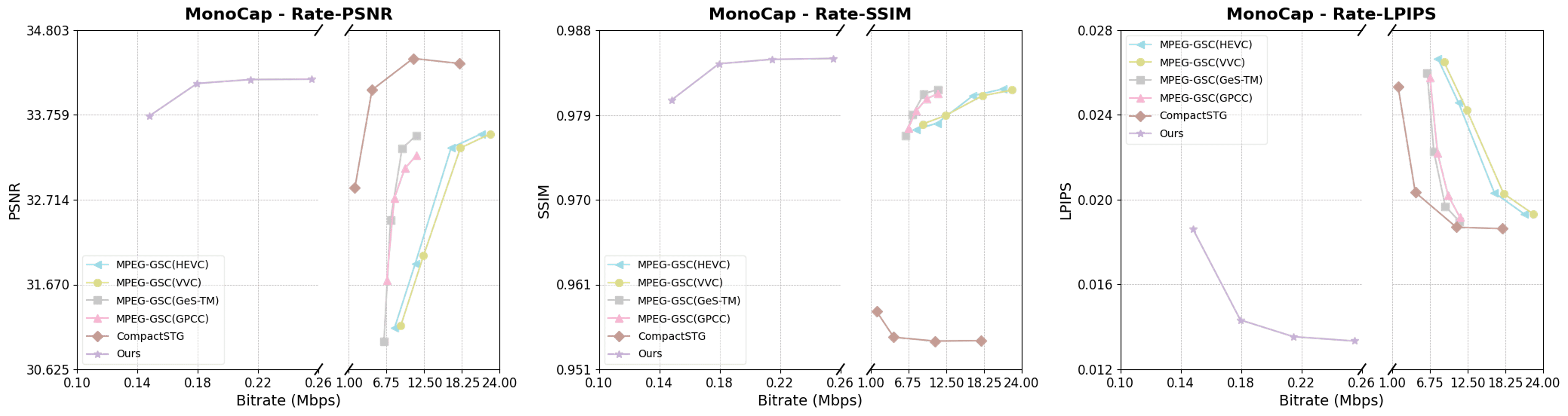}}

\caption{RD performance comparisons in terms of Rate-PSRN, Rate-SSIM and Rate-LPIPS on ZJU-MoCap and MonoCap datasets}
\label{RD}  
\end{figure}

\SubSection{2.4 \ Optimization}
With the canonical-to-target transformation, the canonical 3DGS can be optimized with rendering results on target avatars. The rendering process can be described as, 
\begin{equation}
    \hat{\textbf{I}_{t}^{v}}, \bm{\alpha}_{t}^{v} = splat(\overline{\hat{\mathbf{p}}}_t, \bm{\Sigma}_{t}, \textbf{sh}, \textbf{opa}, v),
\end{equation}
where $\overline{\hat{\mathbf{p}}}_t$ and $\bm{\Sigma}_{t}$ can be obtained from equation(\ref{target_pos}) and equation(\ref{target_cov}), respectively. $\textbf{sh}$ and $\textbf{opa}$ denote the spherical
harmonics coefficients and opacity of the gaussians. $splat$ denotes the splatting process, $v$ denotes the view-point, $\hat{\textbf{I}_{t}^{v}}$ denotes the rendered image, and $\bm{\alpha}_{t}^{v}$ denotes the rendered opacity map. 
Accordingly, the loss function can be defined as,
\begin{equation}
    L = ||\hat{\textbf{I}_{t}^{v}} -\textbf{I}_{t}^{v}||_{1} + \lambda_{1}||\bm{\alpha}_{t}^{v} -\textbf{m}_{t}^{v}||_{2} + \lambda_{2}(1-ssim(\hat{\textbf{I}_{t}^{v}},\textbf{I}_{t}^{v})) + \lambda_{3}lpips(\hat{\textbf{I}_{t}^{v}},\textbf{I}_{t}^{v}),
\end{equation}
where L1 norm, L2 norm, Structural Similarity Index Measure~(SSIM)~\cite{ssim} and Learned Perceptual Image Patch Similarity~(LPIPS)~\cite{lpips} are implemented as loss terms to compared the rendered results with original image $\textbf{I}_{t}^{v}$ and original mask $\textbf{m}_{t}^{v}$. Empirically, $\lambda_{1}$ is set as 0.1 and $\lambda_{2}$, $\lambda_{3}$ are set as 0.01.

\Section{Experimental Results}
\SubSection{3.1 \ Experimental Settings}

~~\quad \textbf{Datasets.} We evalute the proposed framework on two widely used multi-view human video datasets, i.e., ZJU-MoCap~\cite{nerualbody} and MonoCap~\cite{monocap_source, dynacap, deepcap}. For ZJU-MoCap, following practices in~\cite{gauhuman,monocap_source}, six human subjects are selected~(377, 386, 387, 392, 393, 394), where each subject contains multi-view videos from 23 cameras and more than 600 frames for each view. MonoCap has four human subjects with multi-view videos, where two of them have 11 views, two of them have 50 views and each of them has more than 600 frames for each view. For each of these ten sequences, we sample 100 frames with an interval of 5 for both training and testing, and equally number of views are selected for training and testing.

\begin{figure}[t!]
\centering
\subfloat[``my337" of ZJU-MoCap at PSNR of 34dB]{\includegraphics[width=15cm]{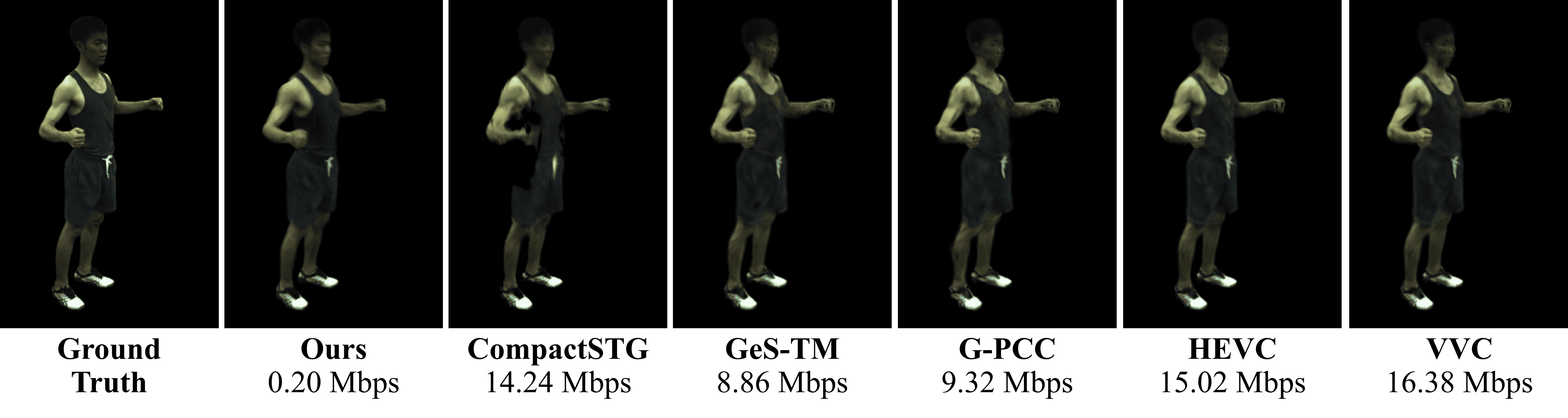}} \\
\subfloat[``lan\_images620\_1300" of MonoCap  at PSNR of 33dB]{\includegraphics[width=15cm]{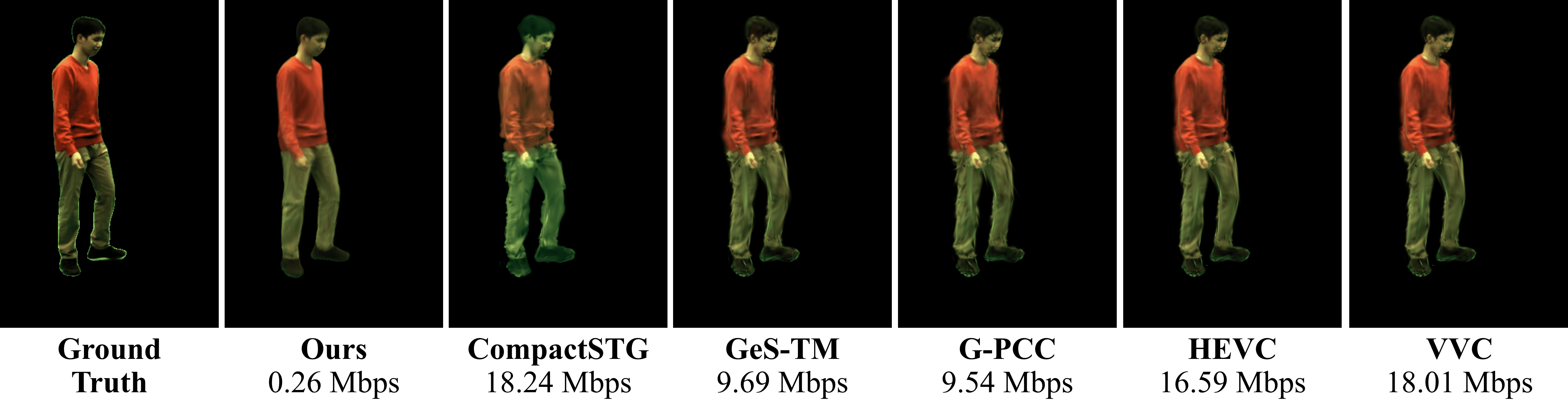}}

\caption{Subjective comparisons on ZJU-MoCap~\cite{nerualbody} and  MonoCap~\cite{monocap_source} dataset at similar quality}
\label{subj}  
\end{figure}
\textbf{Comparison Methods.} We compare the proposed method to both conventional 2D/3D codecs and learning-based dynamic 3DGS compression method. Specifically, for conventional codecs, 3DGS coding anchors from Joint Exploration Experiment 6.2 between WG 4 and WG 7 of The Moving Picture Experts Group~(MPEG)~\cite{gsc-anchor} are adopted, including Point-Cloud-Compression(PCC)-based methods with G-PCC~\cite{gpcc-anchor} and GeS-TM~\cite{ges-tm-anchor}, as well as video-based methods~\cite{video-anchor} with HEVC~\cite{hevc} and VVC~\cite{vvc}. For learning-based dynamic 3DGS compression method, CompactSTG~\cite{CompactSTG} is adopted, where mask-based pruning, network-based color prediction and residual vector quantization are employed for compressing space-time 3DGS~\cite{stgs}. 

\textbf{Implementation Details.} For conventional codecs, we generate frame-by-frame PLY sequences by training every frame as a single multi-view scene with 2000 iterations, and we follow the rate points in~\cite{gpcc-anchor,ges-tm-anchor,video-anchor} for rate control. For CompactSTG, the whole sequence is trained as a dynamic scene with 25000 iterations with its default settings, and we use 4 different pruning coefficients to adjust the compression ratio. 
And for our method, the canonical 3DGS avatar is trained for 25000 iterations, and GeS-TM~\cite{ges-tm-anchor} codec is utilized for canonical avatar compression with 4 different rate points. For bit-rate calculation, we use mega-bits per second~(Mbps) and set Frame-per-Second~(FPS) as 25.
For evaluation metrics, PSNR, SSIM and LPIPS are measured and rate-distortion (RD) curves are used to compare the proposed method with comparison methods.

\SubSection{3.2 \ Evaluation Results}

\begin{figure}[t!]
\centering
\subfloat[``olek\_images0812" of MonoCap]{\includegraphics[width=6.5cm]{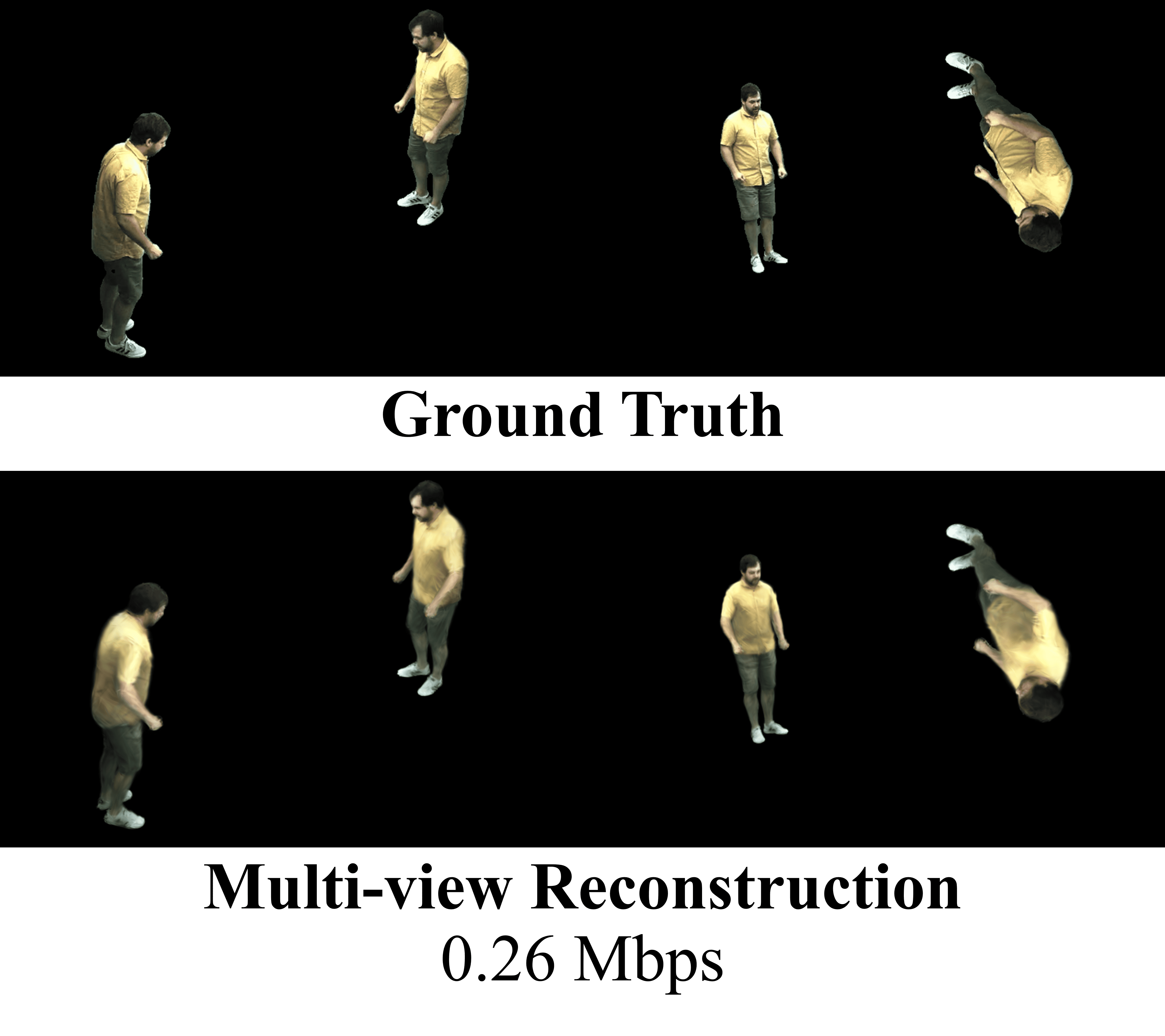}}
\hspace{1cm}
\subfloat[``vlad\_images1011" of MonoCap]{\includegraphics[width=6.5cm]{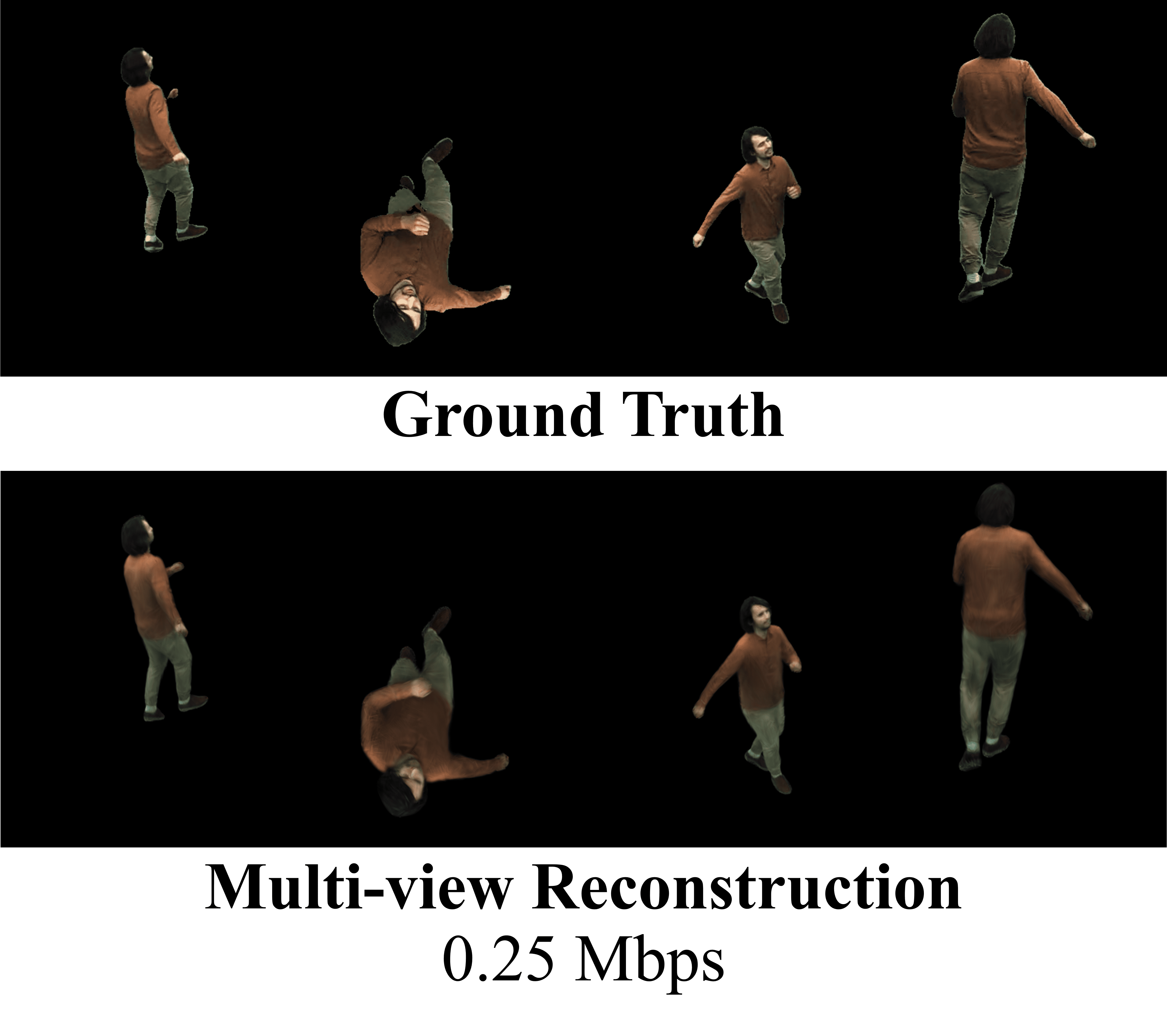}}

\vspace{-0.3cm}
\caption{Multi-view reconstruction results of proposed method}
\vspace{-0.3cm}
\label{multiview}  

\end{figure}

~~\quad \textbf{Rate-Distortion Performance.} The rate-distortion performances in terms of Rate-PSNR, Rate-SSIM and Rate-LPIPS on ZJU-MoCap and MonoCap dataset are shown in Figure~\ref{RD}. The conventional codecs from MPEG are denoted as ``MPEG-GSC(Codec-type)".
It can be seen that the proposed method achieves ultra-low bit-rate of less than 0.2 Mbps on the ZJU-MoCap dataset and less than 0.26 Mbps on the MonoCap dataset, compared to bit-rates exceeding 1 Mbps for comparison methods, demonstrating the effectiveness of leverage highly compact human-prior parameters for temporal modeling. On ZJU-MoCap dataset, conventional codecs exhibit higher-quality upper-bound, which is potentially due to the frame-by-frame training without canonical-to-target transformation. However, the proposed method demonstrates better qualities on Monocap dataset, where human figures are smaller in the scenes with larger global movements, even with lower bit-rate. Overall, the proposed method achieves superior RD performance on both the ZJU-MoCap and MonoCap datasets, while PCC-based methods outperform video-based methods on human avatar sequences. In contrast, learning-based dynamic 3DGS compression methods exhibit less stable performances, particularly failing to perform well on SSIM measurements for the MonoCap dataset.

\textbf{Subjective Quality.} The subjective comparisons of proposed method and comparison methods are shown in Figure~\ref{subj}. Under the similar PSNR measurements, our proposed method can achieve the most visual-pleasing renderings under the lowest bit-rate consumption. Specifically, CompactSTG reconstructions exhibit obvious distortions with large occlusion on ``my377" and color deviation on ``lan\_images620\_1300", while the reconstruction of conventional codecs preserve less detail on human faces and are poorly-rendered on the edges of the bodies.
Besides, the multi-view reconstruction results are shown in Figure~\ref{multiview}, where four different views from 2 sequences of MonoCap are displayed. The proposed method can achieve high-quality rendering on multiple views, demonstrating the high accuracy of target avatar recovery.


\Section{Conclusion}
In this paper, we propose to compress volumetric human video with 3DGS representation in a prior-guided manner. By training a canonical 3DGS avatar and extract human parameters of each timestamp at the encoder side, the appearance and temporal modeling are decomposed for high-efficiency and ultra-low bit-rate transmission. Furthermore, the decoder side is equipped with LBS-based canonical-to-target transformation, which enables both position and shape transform of each 3D gaussian, leading to high-quality target avatar recovering and multi-view rendering. The experimental results demonstrate that the proposed method can achieve superior RD performances compared to both conventional-codec-based 3DGS compression methods and learning-based dynamic 3DGS compression method, shading light on efficient immersive multi-media communication for meta-verse applications.

\Section{References}
\bibliographystyle{IEEEbib}
\bibliography{refs}

\end{document}